# Terahertz-driven Two-Dimensional Mapping for Electron Temporal Profile Measurement

Xie He[1], Jiaqi Zheng[1], Dace Su[1], Jianwei Ying[1], Lufei Liu[2], Hongwen Xuan[2], Jingui Ma[1], Peng Yuan[1], Nicholas H. Matlis[3], Franz X. Kärtner[3,4], Dongfang Zhang[1*], and Liejia Qian[1*]

**Affiliations:**

[1]Key Laboratory for Laser Plasmas (Ministry of Education), School of Physics and Astronomy, Shanghai Jiao Tong University, 200240 Shanghai, China

[2]GBA Branch of Aerospace Information Research Institute, Chinese Academy of Sciences, 510700 Guangzhou, China

[3]Center for Free-Electron Laser Science, Deutsches Elektronen-Synchrotron, Notkestrasse 85, 22607 Hamburg, Germany

[4]Department of Physics and The Hamburg Centre for Ultrafast Imaging, Universität Hamburg, Luruper Chaussee 149, 22761 Hamburg, Germany

* Corresponding author: dongfangzhang@sjtu.edu.cn; qianlj19@sjtu.edu.cn

**Abstract**: The precision measurement of real-time electron temporal profiles is crucial for advancing electron and X-ray devices used in ultrafast imaging and spectroscopy. While high temporal resolution and large temporal window can be achieved separately using different technologies, real-time measurement enabling simultaneous high resolution and large window remains challenging. Here, we present the first THz-driven sampling electron oscilloscope capable of measuring electron pulses with high temporal resolution and a scalable, large temporal window simultaneously. The transient THz electric field induces temporal electron streaking in the vertical axis, while extended interaction along the horizontal axis leads to a propagation-induced time delay, enabling electron beam sampling with sub-cycle THz wave. This allows real-time femtosecond electron measurement with a tens-of-picosecond window, surpassing previous THz-based techniques by an order of magnitude. The measurement capability is further enhanced through projection imaging, deflection cavity tilting, and shorted antenna utilization, resulting in signal spatial magnification, extended temporal window, and increased field strength. The technique holds promise for a wide range of applications and opens new opportunities in ultrafast science and accelerator technologies.

## Introduction

Real-time temporal profile measurements are essential for advancing applications in ultrafast electron diffraction[1,2], microscopy[3], and free electron laser sources[4]. Various conceptual and experimental approaches have been explored for characterizing electron pulse temporal profiles spanning from the picosecond to the femtosecond regime[5-16]. The streak cameras have been the conventional solution for real-time measurement of an electron beam. It converts the temporal profiles to spatial profiles by pulling electrons with a sweep voltage, enabling the electron bunch length measurement from the keV and GeV electron energy with fs resolution. Radio-frequency (rf) cavities-based streak cameras have been widely used in accelerator facilities providing a temporal window of a few tens of picosecond (ps)[17-19]. However, phase jitter issues inherent to RF cavities typically limits its system temporal resolution in accumulation mode to tens of fs[20,21]. The considerable size of the RF cavity limits its application in precision measurements, particularly in experiments like ultrafast electron diffraction, where accuracy at the sample location is crucial. Furthermore, for low-energy electron sources, its substantial dimensions will lead to bunch lengthening due to drift and space charge effects, consequently diminishing measurement precision. While large streaking gradients on the order of μrad/as can be achieved using near-infrared (NIR) lasers, enabling attosecond resolution, these methods are typically limited to a temporal window of only a few femtoseconds[22,23].

Recently, with the development of strong field THz[24-26], there has been significant interest in the use of THz-based streaking technologies to measure the temporal profile of the electron beam[8-11]. THz waves inside a resonator can provide a field gradient on the order of hundreds of MV/m, enabling a temporal resolution of a few fs with millimeter scale devices. The THz-based streaking in a slit antenna also provides a solution for characterization of the arrival time of electron beams by encoding the time of arrival of the electron beam onto the spatial beam position on the detector, a capability particularly valuable for RF-powered MeV ultrafast electron diffraction applications[10,11]. However, the angular streaking limits its temporal window to about a quarter of the THz wave, normally in the sub-ps regime[9,11]. Such slit antenna has also been extended to measure the optical waveform by injecting the optical pulse transversely and interact with a short electron beam[27]. Although a circularly polarized THz wave with a dielectric tube is proposed to extend the dynamic range[28], the temporal window is still constrained by the length of the single-cycle THz wave. Other solutions such as electro-optic sampling[29,30] and transition radiation[31-33], also exhibit limited temporal windows while at the same time requires high bunch charge (pC) and energy (MeV) to produce a decent signal. Cross-correlation of the electron bunch with a laser pulse based on ponderomotive scattering or strong field deflection in principle can significantly extend the temporal window and increase resolution[7,14,34]. However, their ultimate resolution is limited by the duration of the pump laser and cannot support real-time measurements. While obtaining high temporal resolutions and large temporal windows is conceivable separately using different methods, simultaneous real-time measurement seems out of reach with current technologies.

Here, we demonstrate a THz-driven electron oscilloscope capable of measuring the temporal information of an electron beam with simultaneous high temporal resolution and a scalable large temporal window. This scheme exploits a laser machined antenna with a slit, producing ultra-wide coupling bandwidth and high directivity. The field is spatially confined to a sub-wavelength regime, enabling a record streaking speed of > 200 µrad/fs. The transient electric field induces streaking of the electron beam along the vertical axis (y axis), imprinting the electron temporal information into space with fs temporal resolution. While extension in the horizontal direction (x axis) leads to spatial-dependent time delay that enables sampling of the electron beam with the sub-cycle THz wave, hence, encoding the electron beam temporal information into the x axis with a moderate temporal resolution but a larger temporal window. Furthermore, we demonstrate the extension of this technique through projection imaging, deflection cavity tilting, and shorted antenna utilization, resulting in signal spatial magnification, extended temporal window, and enhanced field strength. Additionally, the system features straightforward fabrication, with the antenna manufactured using standard laser cutting technology on metallic thin plates.

## Results

### Concept and implementation

The experimental setup Fig. 1a consists of a single-cycle THz generation stage, a direct-current (DC) gun, and an ultraviolet (UV) pulse. The electrons generated in DC gun are accelerated by the DC field to 40 keV and measured by a microchannel plate (MCP) detector. THz pulses were coupled into the device transversely by a THz lens with its polarization vertically along the y axis.

To achieve a high THz field gradient, an exponentially shaped Vivaldi antenna[35,36] is designed for coupling the THz wave. It enables ultra-wide band coupling and focusing of the THz wave into the center slit, which ensures single-cycle propagation along the slit and sub-wavelength concentration of the THz wave. The Vivaldi antenna provides an overall 18 times field enhancement at the center (Fig. 2a, b), which is ~8 times higher than injection from the front using a simple slit, as employed in other THz-based streak cameras[11] with a 100 μm slit gap. Meanwhile, front injection of THz wave typically resulted in larger dispersion, leading to the development of a long THz wave (Fig. 2b). A horn antenna is also designed to further confine the field and hold the antenna inside the vacuum chamber (Fig. 1a).

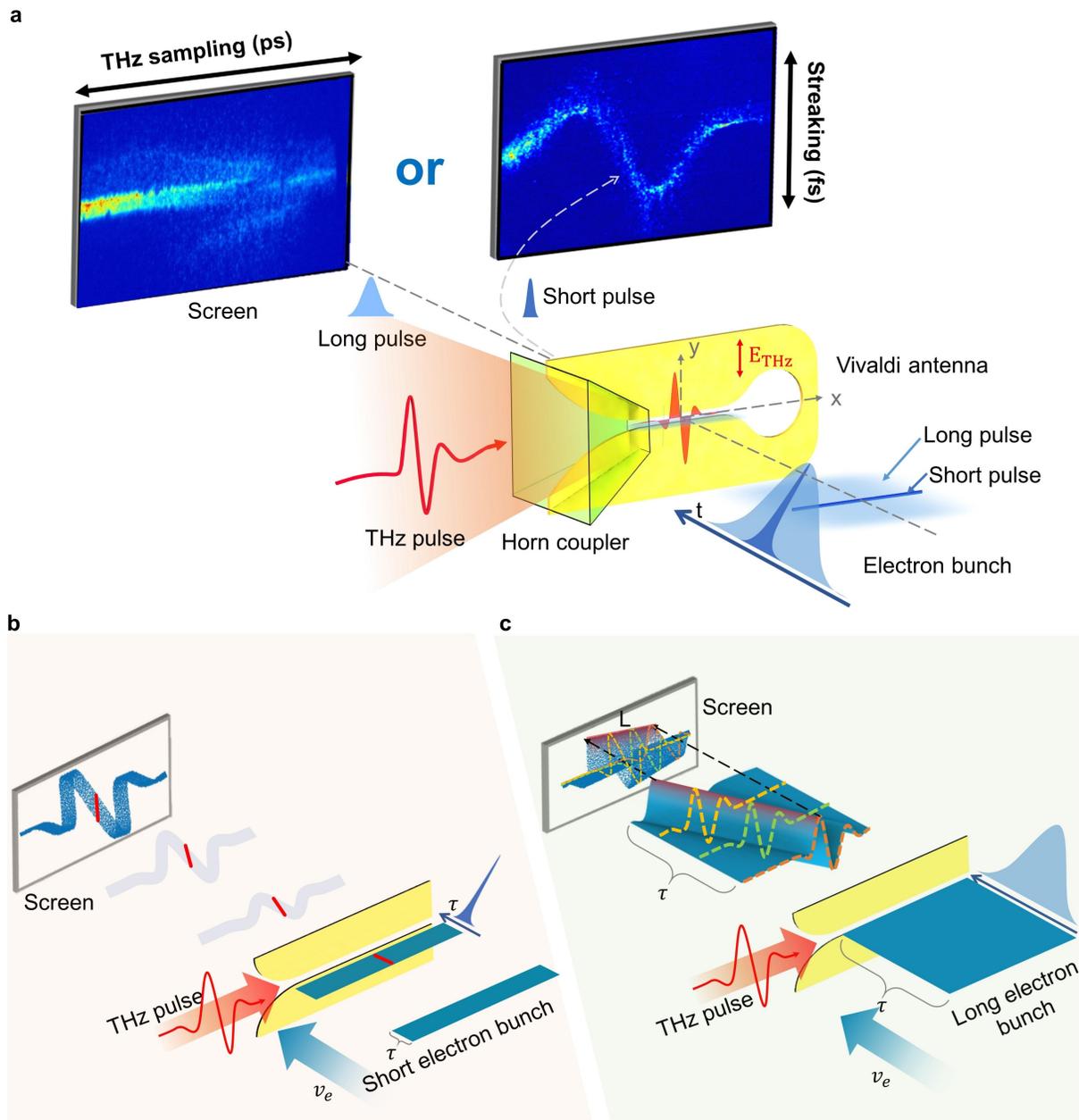

**Fig. 1 Proof-of-concept for real-time electron bunch length measurement. a** A photo-triggered DC gun generates pulsed electron beam. A vertically polarized THz wave is focused onto the antenna through a THz lens. The electron beam is measured by a MCP detector. The THz electric field deflects the electron beam in the vertical (y) direction. The THz-electron interaction is extended in the horizontal direction (x) along the center slit, enabling sampling of the electron beam with sub-cycle THz wave. **b, c** Schematic illustration of THz-electron interaction for both short (**b**) and long (**c**) electron bunches. For the long bunch, the interaction resembles that with a series of short bunches, with their deflection signals overlapping on the detector and shifted horizontally along the x-axis.

An electron beam, elongated spatially along the x-axis and confined along the y-axis, is injected perpendicularly to the slit region of the cavity (Fig. 1). During the THz-electron interaction, the electrons experience the electric field for deflection according to the Lorentz force law $\vec{F} = e(\vec{E} + \vec{v} \times \vec{B})$, where $e$ is the electron charge, $\vec{E}$ is the electric field that is perpendicular to the electron propagation, and $\vec{B}$ is the magnetic field. The timing between the electron and THz is controlled by motorized stages for each relative IR pump beam generating the relative UV photoexcitation laser and THz wave. When the electrons sweep the zero-crossing cycle of the THz $\vec{E}$ field (red marked in Fig. 1b), they will experience a strong deflection as a function of time enabling the measurement of the temporal bunch profiles by mapping them onto the spatial dimension of the MCP detector (Fig. 1b). Consequently, the pulse duration from a short electron bunch can be derived. Such streaking process has also been used by Baek et al., to measure the optical waveform with a short electron beam[27]. While the electron pulse duration exceeds a quarter of the THz wave, the vertical deflecting force no longer increases linearly. The sine (or cosine) shape of the THz wave introduces a nonlinear term to the position-energy relationship within the bunch, necessitating consideration of the curvature of the THz waveform. When the electron bunch is about one-half of the THz wave, the streaked electron beam will no longer extend along the y axis (Fig. 2e, f). Consequently, the bunch length cannot be determined through the vertically deflected electron beam on the detector.

As the THz wave is injected at a 90-degree angle relative to the electron beam, maximum deflection is achieved when the electron beam coincides with the crest of the THz wave. When the electron beam exceeds one-half of the THz wavelength, scanning the relative THz-electron delay enables sampling of electron temporal information through the maximally deflected electron beam. However, obtaining sufficient information to retrieve the electron temporal information requires a delay scan. By extending the electron beam and the slit lateral size along the horizontal direction (x coordinate in Fig. 1c), the sub-cycle THz-driven electron sampling can be imprinted into space. As shown in Fig. 1c, the THz samples different temporal components of the electron beam along a triangular trajectory in the spatially extended electron beam (see Materials and methods for more information). The deflected electron beam along the x axis on the detector directly indicates the electron pulse duration (Red dashed squares in Fig. 2e, f) according to $\tau = L/c$, where $L$ is the lateral length marked in Fig. 2e, f and Fig. 1c, which can be calibrated by the lateral size of the main beam ($L_0$) that goes through the slit without THz deflection. The temporal resolution is determined by the length of the positive (or negative) sub-cycle of the THz wave. The Vivaldi antenna ensures ultra-wide band propagation, thereby preserving a short THz pulse, thus ensuring the temporal resolution for cross-correlation. The temporal window is determined by the electron beam size along the x axis. Due to the attenuation of the THz field as it propagates through the slit, electrons at the front of the bunch interact with the THz pulses earlier and experience a greater deflection force compared to those at the rear of the bunch, causing the red dashed square component to appear slightly slanted (Figs. 2e and 2f).

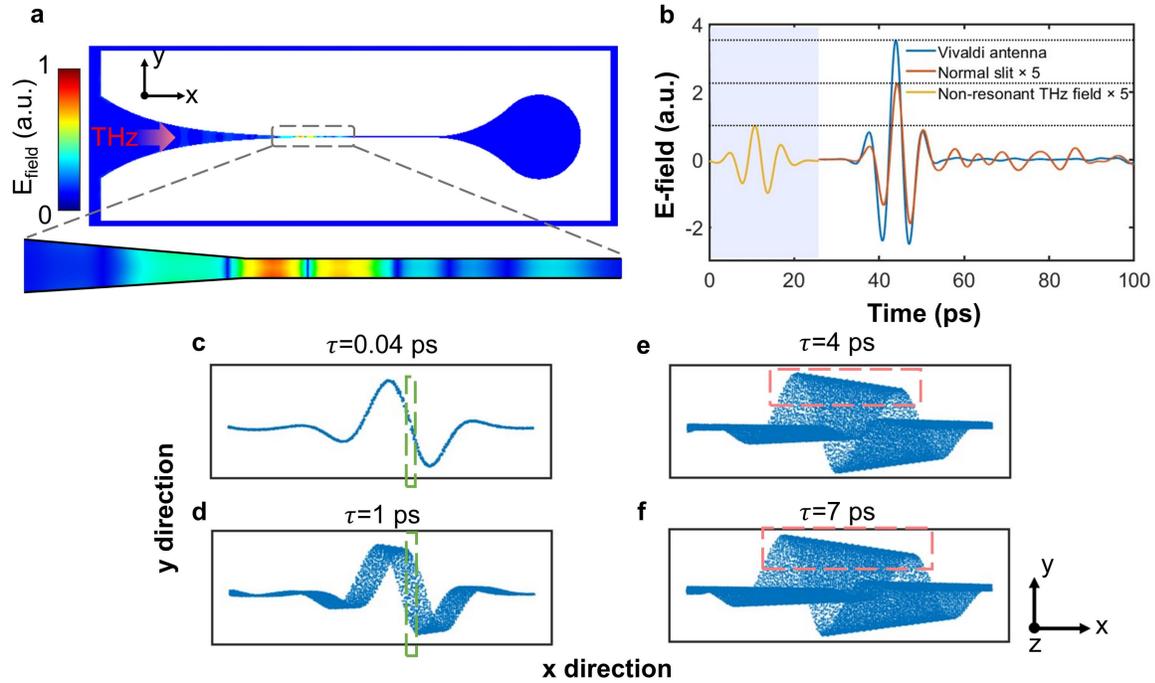

**Fig. 2 Proof-of-concept of the electron temporal profile measurement. a** Vivaldi antenna with a slit and its simulated lateral field distribution with a single-cycle THz wave injected from the side. **b** Simulated THz field at the center interaction region for a slit with Vivaldi antenna (blue), a slit with front incident THz wave (red), and initial injected THz wave (yellow). The label 'x5' indicates a scale factor applied for visualization purposes. **c-f** Simulated electron transverse beam profile at the detector position for initial electron pulses with durations of 0.04 ps, 1 ps, 4 ps, and 7 ps (FWHM). The green and red dashed squares in panels (**c**)-(**f**) indicate the regions used to derive the electron pulse duration using streaking and cross-correlation, respectively.

The THz-electron interaction inside the slit generates a two-dimensional map on the detector, with streaking along the vertical (y) axis and extended sub-cycle THz sampling along the horizontal (x) direction. Direct injection of a large electron beam into the slit will reduce the number of electrons passing through the slit, hence, a quadrupole magnet is preferred, which can focus the electron beam in one dimension and defocus it in the other dimension, generating an elliptical beam suitable for a larger temporal window. Additionally, the antenna is designed with a water droplet stub to prevent back reflection of the THz wave.

**High-resolution and large temporal window measurement**

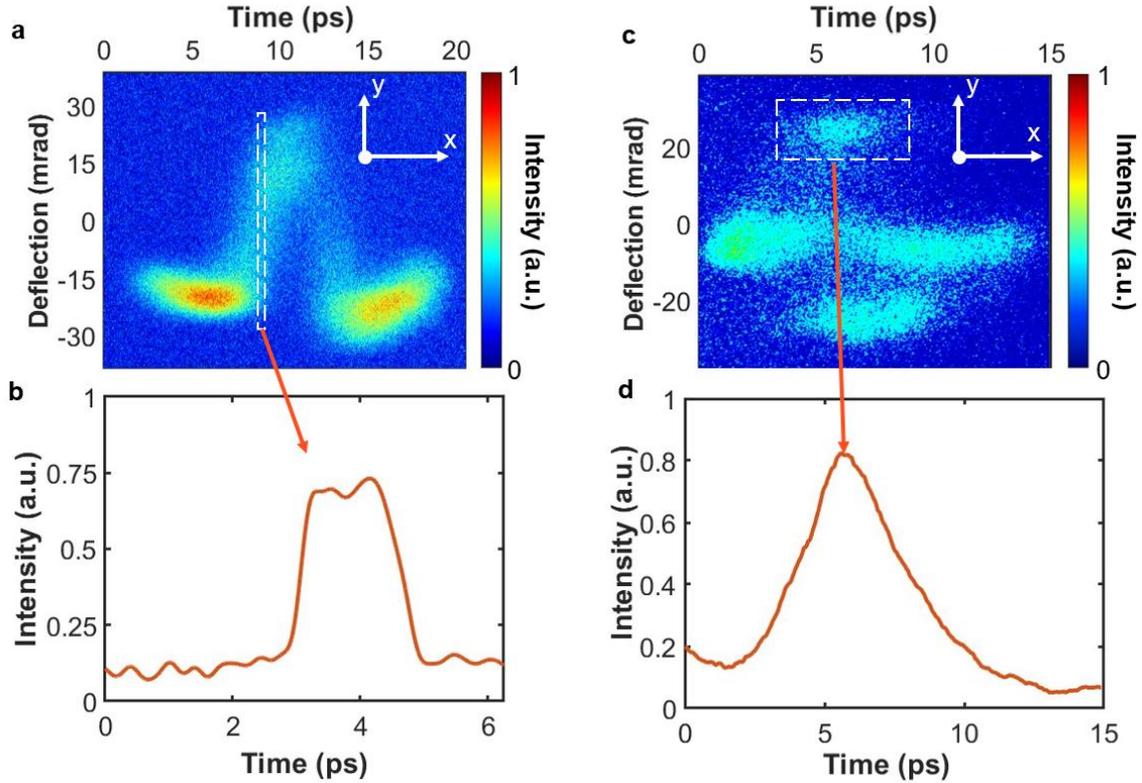

**Fig. 3 High-resolution and large temporal window electron temporal profile measurement.** Raw image recorded on the MCP detector for a short (**a**) and long (**c**) electron bunch. **b**, **d** The corresponding retrieved temporal profiles of the electron pulses reveal FWHM pulse durations of (**b**) 1.6 ps and (**d**) 4.1 ps, respectively. The THz wave with a carrier-envelope phase of zero degrees (Fig. 3a) exhibits one dominant positive peak and two smaller negative peaks. During cross-correlation, a shorter probe provides higher temporal resolution. Consequently, the top component is used to retrieve the temporal profile.

Figure 3 presents experimental results obtained from short and long electron beams emitted from the DC gun. In the case of a short electron beam, a distinct sine wave is observed, representing the profile of the THz pulse with a carrier-envelope-phase of zero degrees. A slice of the zero-crossing component (dash marked square in Fig. 3a) shows the streaking in the vertical direction, with a retrieved pulse duration of ~1.6 ps (FWHM) (Fig. 3b). As the duration of the electron pulse increases, there is a noticeable increase in the horizontal component (dashed square in Fig. 3c). However, the top and bottom horizontal components do not have equal lengths. This discrepancy is primarily due to the THz carrier-envelope-phase being at zero degrees, resulting in one positive peak and two small negative peaks, as also indicated in Fig. 3a. Clearly, the single positive peak provides higher resolution. The bunch length should be determined from the shorter (top) component, indicating an electron pulse duration of 4.1 ps (FWHM).

The temporal window of the aforementioned scheme is primarily determined by the lateral dimensions of the electron beam. Increasing the beam diameter will lower the electron density, thereby reducing the signal intensity. To further extend the temporal window without sacrificing signal intensity, the deflection cavity can be tilted at an angle θ. This increases the travel distance of the electron beam inside the device (Fig. 4a) and the effective interaction length, hence the maximal bunch length:

$$\tau_{max} \approx \frac{L_0 * (v_e + v_{THz} * sin\theta)}{v_{THz} * v_e * cos\theta}$$

where $L_0$ is the lateral size of the electron beam at the slit position, $v_e$ is the electron velocity, and $v_{THz}$ is the THz velocity. Figure 4b illustrates simulated electron bunches at the detector position, indicating an approximately 3.8-fold increase in temporal window when the cavity is tilted by 40°. Figure 4c shows the amplification of the temporal window in the tilted cavity with different electron energy and tilted angle. Figure 4d shows the simulated temporal window with an initial electron energy of 30 keV. As observed, the temporal window is enlarged with the increase of the tilt angle and decrease of the electron energy (speed). Figures 4e, f depicts the measured bunch length on the MCP and its retrieved bunch length with θ=40°. For the same electron bunch, the measured signal for the untilted configuration nearly fills the entire x-axis with a temporal window of ~10 ps. Tilting the cavity at an angle of around 40 degrees extends the time window to nearly 40 ps. An electron pulse duration of approximately 6.5 ps is derived from both measurements, demonstrating the consistency and validity of the time-window stretching technique. However, while tilting the cavity extends the temporal window, it also increases the equivalent time interval per pixel, thereby reducing the temporal resolution. Further extension of the temporal window would require an electron beam with a larger horizontal size, leading to a decreased signal-to-noise ratio on the detector.

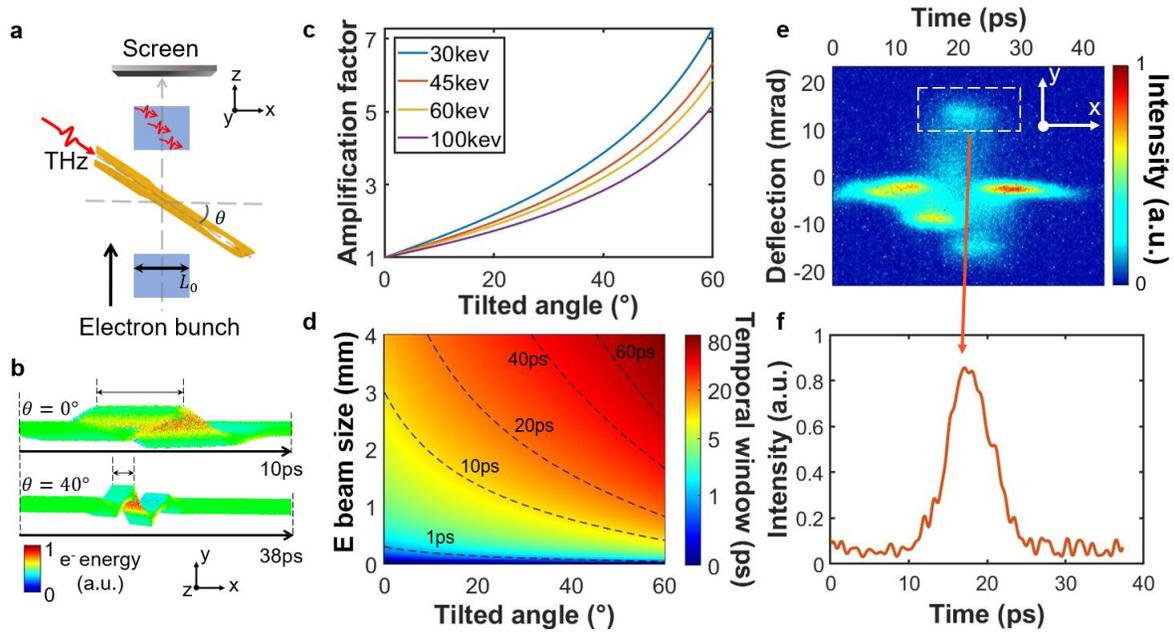

**Fig. 4 Feasibility of extended temporal window with tilted cavity. a** Schematic of the measurement with a tilted antenna. **b** Front views of the simulated energy modulations of the electron bunch induced by the cross-correlation inside the slit when θ=40° and 0°. **c** Temporal window amplification factor in the tilted cavity compared to the untilted situation. **d** Simulated temporal window versus electron beam size

and tilted angle with an initial electron energy of 30 keV. **e, f** Raw images recorded on the MCP detector and the retrieved electron temporal profile (**f**) with a FWHM pulse duration of 6.5 ps. The extended temporal window is achieved with a reduced resolution due to increased pixel time intervals.

**Spatial signal magnification with projection imaging**

The previous demonstration (Fig. 3) was performed with a nearly parallel electron beam with a low divergence angle. To achieve a larger temporal window, an elliptical beam is desired. A single quadrupole magnet, capable of focusing the electron beam in one dimension and defocusing in the other dimension, is ideal for such an application. It also facilitates projection imaging, which spatially magnifies the signal on the detector with a ratio of ($\frac{l_1+l_2}{l_1}$) (Fig. 5a). However, the time difference induced by the divergence angle ($\alpha$) along the antenna slit should be considered. As can be seen from Fig. 5a, compared with the straight beam, the arrival time of an electron beam with a divergence angle of $\alpha$ will experience a longer travel time, resulting in a time delay: $\Delta t = \frac{l_1+l_2}{v_e}\left(\frac{1}{\cos\alpha} - 1\right)$, where $v_e$ is the electron velocity, $l_1$ is the quadrupole to device distance, and $l_2$ is the antenna to MCP detector distance. This timestamp should be considered while retrieving the electron bunch length from the lateral sub-cycle sampling.

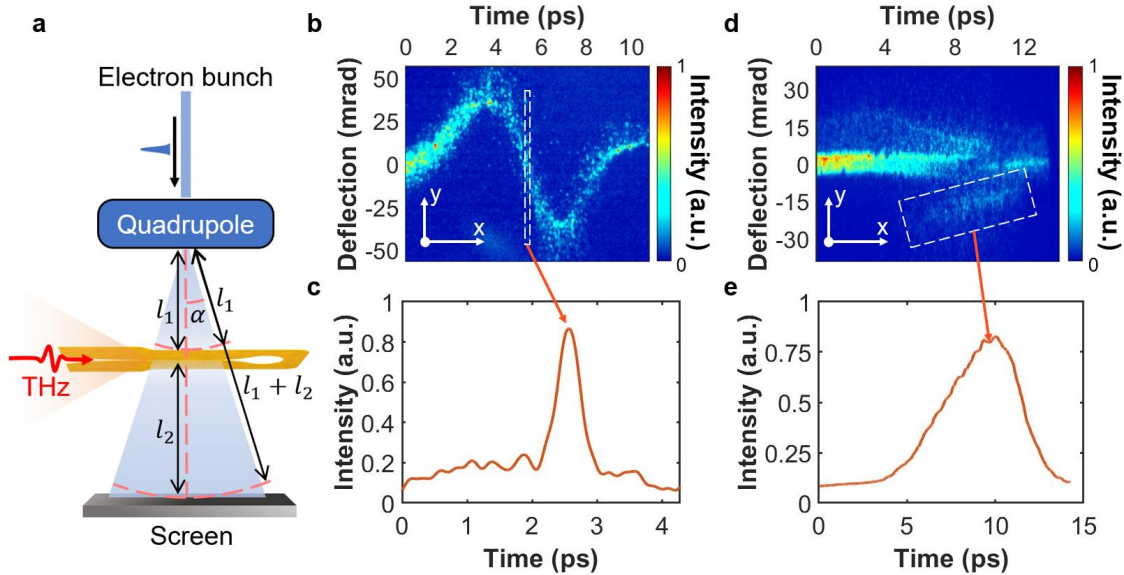

**Fig. 5 Spatial signal magnification with projection imaging. a** Schematic of projection imaging with a quadrupole magnet that focuses in the y axis and defocuses in the x axis. **b**, **d** Raw data recorded on the MCP detector with a short (**b**) and long (**d**) electron beam measured with projection imaging with a solenoid lens. **c, e** The corresponding retrieved temporal profiles of the electron pulses with a pulse duration of (**c**) 344 fs and (**e**) 5.4 ps, respectively. The THz wave with a carrier-envelope phase of 90 degrees (Fig. 5b) has one positive peak and one negative peak. During cross-correlation, both the top and bottom components provide a similar temporal profile.

Due to mechanical constraints in the current DC gun chamber, a quadrupole cannot be used in the DC gun. To demonstrate the concept, we employed projection imaging with a solenoid lens to obtain a comparable imaging signal. However, this approach results in significant electron loss because the magnification is two-dimensional, while the slit selectively captures only one dimension of the signal with only a small portion, leading to a reduced signal-to-noise ratio. Benefiting from the projection imaging, the signal is spatially magnified by a factor of 20, ensuring more pixels occupied on the detector. This spatial magnification improves the resolution and accuracy of the retrieved signal compared to the case without magnification. A sharper signal is achieved for two different cases demonstrating short (344 fs) (Fig. 5b, c) and long (5.4 ps) (Fig. 5d, e) electron pulse durations in this experiment. In the case of the longer electron beam, the two transverse components have almost the same length, which is different from the above measurement (Fig. 3c). This is mainly because the driving THz is at a 90-degree carrier-envelope-phase (Fig. 5b), and the positive and negative components of the single-cycle THz wave have the same duration, which determines the temporal resolution of cross-correlation.

**High gradient streaking**

For THz-based streaking of an electron beam, a deflection of about a few milliradians can already provide few fs time resolution[9-11]. As the THz wave is injected from one side, higher gradient streaking can be achieved by a shorted antenna (Fig. 6a), which generates a superposition of forward and reflected THz fields. With a THz energy of ~0.8 µJ coupled into the device, the maximum streaking field gradient reaches about 250 MV/m, producing a deflection angle of 214 mrad and a streaking speed of 200 µrad/fs as depicted in Fig. 6b. These results represent a new record in THz-based streaking gradient, an order of magnitude higher than previous demonstrations[9-11]. In this shorted antenna, THz sampling along the x axis is not useful anymore as the electron beam will interact with both the forward and reflected THz wave. This design primarily aims for a higher field gradient, thereby achieving better temporal resolution through streaking along the y axis. Although the current deflection angle (proportional to the streaking speed) is two orders of magnitude greater than previous results for electron bunch length[10,11] or THz waveform measurements[27] in the slit, the resolution of the measurement, related to the finite size of the unstreaked beams on the detector (328 µm), is evaluated to be 9 fs. The current resolution is primarily constrained by the size of the unstreaked beam, which is more than 10 times higher than in previous demonstrations[9]. A natural next step will be reducing the vertical beam size of the initial electron beam, which holds the potential to provide sub-femtosecond temporal resolution. The device exhibits high stability and repeatability under a field gradient of 250 MV/m over several hours of operation. For example, it was used to measure the arrival time of the electron beam, showing a timing jitter of approximately 3 fs (rms) over one hour of measurement.

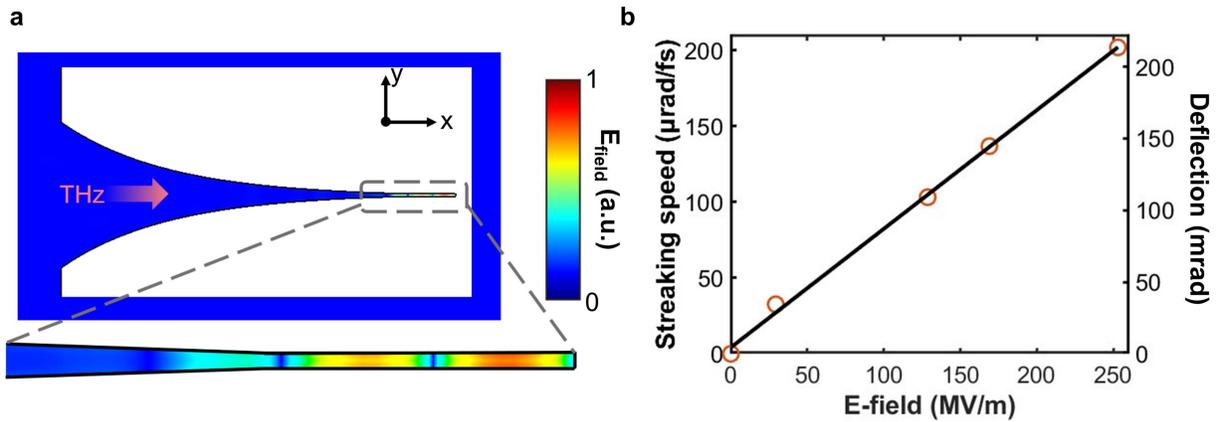

**Fig. 6 High gradient streaking. a** Simulated field distribution of the device with a shorted antenna. **b** Experimental results (red circles) of streaking speed and deflection angle as a function of the input THz field strength. The corresponding electric field is determined by matching the measured deflection angle with simulation results. The relative deflection angle is shown on the right vertical axis.

## Discussion

We have demonstrated a simple and robust technique for characterizing the temporal information of the electron beam with simultaneous high temporal resolution and large temporal window, utilizing a THz powered cavity. The exponentially shaped Vivaldi antenna enables ultra-wide band coupling and sub-wavelength concentration. The electron temporal information is encoded from time to two-dimensional space, which are simultaneously imprinted onto the two orthogonal coordinates (x, y) on the MCP detector, facilitating high temporal resolution and a large temporal window. Furthermore, the field gradient and resolution can be further enhanced by employing a shorted antenna, which creates a superposition of forward and reflected THz fields, resulting in a record streaking speed of 200 μrad/fs. Additionally, we have expanded the capabilities of this technique through the implementation of projection imaging and tilting of the deflection cavity for further signal spatial magnification and extension of the temporal window. This showcase of a THz-driven electron oscilloscope represents a significant advancement in beam optimization for ultrafast science. More broadly, this technique would be advantageous for many accelerator applications, such as ultrafast electron diffraction and laser-driven accelerators, that demand high timing stability and short bunch lengths.

## Materials and methods

### Experimental details

In our experiment, a 1.4 ps, totally 100 mJ, 1030 nm Thin-disk YAG laser operating at 1 kHz is used and about 10% laser energy (10 mJ) is used. THz pulses are generated by the well-established tilted pulse-front method in Lithium Niobate crystals ($LiNbO_3$) with a center frequency of 0.15 THz, 5 μJ pulse energy. UV pulses for photoemission were generated by two successive second harmonic generation stages, and then electron pulses are accelerated by DC field.

### Design of the Vivaldi antennas

The stainless-steel Vivaldi antenna in Fig. 2a has a 100 μm slit gap, 10mm slit length and a thickness of 50 μm, while the shorted antenna in Fig. 6a has a 2mm slit length. Horn-like structures with a port size of 9.6*10 mm and a total length of 14 mm are used in both antennas for better confinement of the THz field and holding the antenna inside the vacuum chamber.

### Analysis of Streaking and Cross-correlation

The analysis of streaking for a short electron bunch is similar to prior studies[8-11], where the streaking deflectogram is measured through a THz-electron delay scan. In the current configuration, the transverse injection of the THz wave introduces a time delay along the x-axis, enabling real-time acquisition of the deflectogram. As depicted in Fig. b, when the electron beam is short, each segment of the electron beam along the x-axis sweeps only a small portion of the THz wave, resulting in a near sine or cosine wave on the detector. Near the zero-crossing (marked in red in Fig. 1b), the front and rear components of the electron beam along the propagation direction (z) experience strong deflection as a function of time due to the THz field. As the electron pulses are deflected in the y-direction by the THz field, their temporal profiles are transformed into spatial profiles (Fig. 1b). This enables the measurement of the temporal bunch profiles by measuring the spatial dimension of the MCP detector, analogous to the typical streak camera technique (Detailed simulation of the top view of the energy modulation can be found in Fig. S1).

Concerning the cross-correlation, Fig. 7 shows a schematic of the THz-electron interaction with a long electron pulse (blue rectangle). When $t = t_0$, as shown in Fig. 7a, neither the THz pulse nor the electron bunch has reached the interaction region in the antenna, therefore, no deflection occurs (Fig. 7a front view). As the electron bunch enters the interaction region and begins to interact with the THz wave (Fig. 7b $t = t_1$), the transverse THz wave injection introduces a time delay along the x axis relative to the components of the electrons along the propagation direction (z). As the interaction continues, with the propagation of both THz and electron beam, more electrons are sampled and deflected by the THz wave. This process can be visualized as the THz wave interacting sequentially with a series of short electron pulses, each with a time shift $\Delta t$ relative to the THz wave. The result is an extension of the deflected electron beam along the $x$ axis (Figs. 7c and 7d), which corresponds to the cross-correlation of the electron bunch $I(t)$ and the sampling THz pulse $S(t)$. The resulting cross-correlation function is given by: $C(\Delta t) = \int I(t)S(t + \Delta t)dt$, representing the deflected electron distribution along the $x$ axis. In Fig. 7d, the white dashed line within the electron bunch (of length $L_{inter}$) indicates the portion of the electron beam that is deflected by the THz wave. The transverse ($x$) component of $L_{inter}$, denoted as $L$, can be expressed as: $L = v_{THz} * t$, where $v_{THz}$ is the velocity of THz wave in vacuum, and $t$ is the interaction time between the THz and the electron bunch, which equals the electron pulse duration $\tau$. Therefore, the observed length $L$ on the detector, which is a direct projection of the interaction length, is linearly proportional to the bunch length $\tau$. This relationship allows the electron pulse duration to be calculated as: $\tau = L/v_{THz}$.

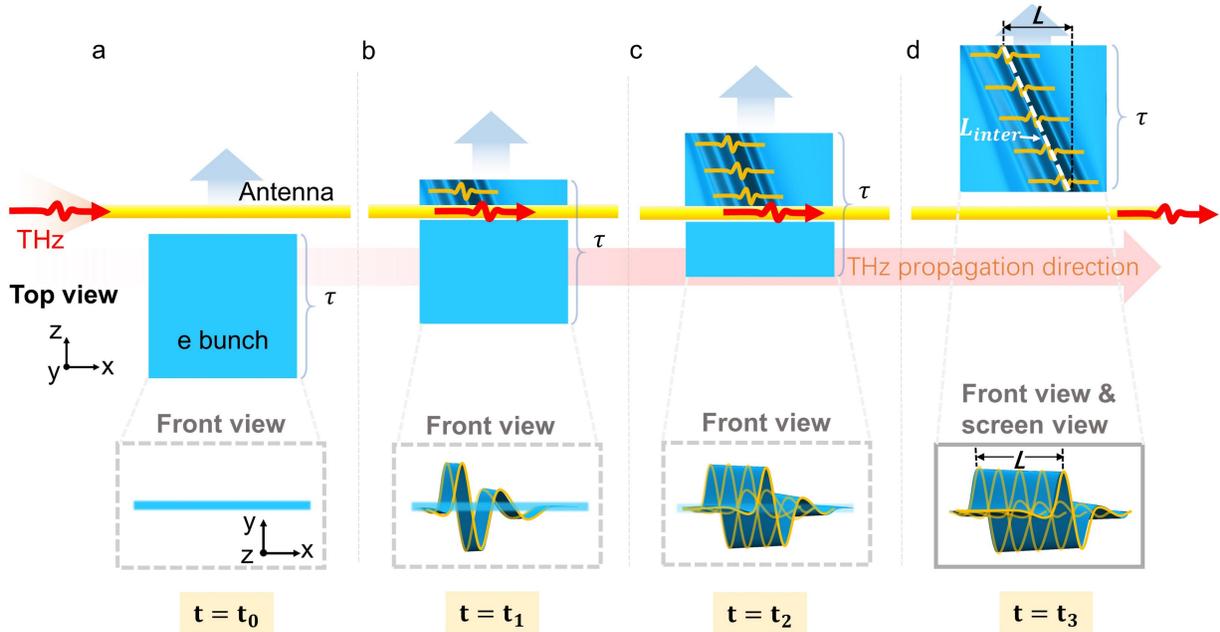

**Fig. 7 THz-electron cross-correlation process with a long electron pulse. a - d** Top views and the corresponding front views (bottom of Fig. 7) of the instantaneous electron pulse at different times during the THz-electron interaction ($t = t_0, t_1, t_2, t_3$). The red curve represents the THz wave, while the yellow

curves indicate the deflected components of the electron beam. **d** The front view and detector (screen view) of the electron beam, where L represents the projection of the interaction length ($L_{inter}$) on the x axis.

# References

1. Ischenko, A. A., Weber, P. M. & Miller, R. J. D. Capturing chemistry in action with electrons: realization of atomically resolved reaction dynamics. *Chemical Reviews* **117**, 11066-11124 (2017).
2. Filippetto, D. *et al.* Ultrafast electron diffraction: Visualizing dynamic states of matter. *Reviews of Modern Physics* **94**, 045004 (2022).
3. Zewail, A. H. 4D ultrafast electron diffraction, crystallography, and microscopy. *Annual Review of Physical Chemistry* **57**, 65-103 (2006).
4. Bressler, C. & Chergui, M. Ultrafast X-ray absorption spectroscopy. *Chemical Reviews* **104**, 1781-1812 (2004).
5. Fabiańska, J., Kassier, G. & Feurer, T. Split ring resonator based THz-driven electron streak camera featuring femtosecond resolution. *Scientific Reports* **4**, 5645 (2014).
6. Musumeci, P. *et al.* Capturing ultrafast structural evolutions with a single pulse of MeV electrons: Radio frequency streak camera based electron diffraction. *Journal of Applied Physics* **108**, 114513 (2010).
7. Hebeisen, C. T. *et al.* Grating enhanced ponderomotive scattering for visualization and full characterization of femtosecond electron pulses. *Optics Express* **16**, 3334 (2008).
8. Zhang, D. *et al.* Segmented terahertz electron accelerator and manipulator (STEAM). *Nature Photonics* **12**, 336-342 (2018).
9. Kealhofer, C. *et al.* All-optical control and metrology of electron pulses. *Science* **352**, 429-433 (2016).
10. Li, R. K. *et al.* Terahertz-based subfemtosecond metrology of relativistic electron beams. *Physical Review Accelerators and Beams* **22**, 012803 (2019).
11. Zhao, L. *et al.* Terahertz Streaking of Few-Femtosecond Relativistic Electron Beams. *Physical Review X* **8**, 021061 (2018).
12. Kirchner, F. O., Gliserin, A., Krausz, F. & Baum, P. Laser streaking of free electrons at 25 keV. *Nature Photonics* **8**, 52-57 (2013).
13. Kassier, G. H. *et al.* A compact streak camera for 150 fs time resolved measurement of bright pulses in ultrafast electron diffraction. *Review of Scientific Instruments* **81**, 105103 (2010).
14. Chatelain, R. P., Morrison, V. R., Godbout, C. & Siwick, B. J. Ultrafast electron diffraction with radio-frequency compressed electron pulses. *Applied Physics Letters* **101**, 081901 (2012).
15. Tsarev, M., Ryabov, A. & Baum, P. Measurement of Temporal Coherence of Free Electrons by Time-Domain Electron Interferometry. *Phys Rev Lett* **127**, 165501 (2021).
16. Yajima, W. *et al.* Streaking of a Picosecond Electron Pulse with a Weak Terahertz Pulse. *ACS Photonics* **10**, 116-124 (2023).
17. Alesini, D. *et al.* RF deflector design and measurements for the longitudinal and transverse phase space characterization at SPARC. *Nuclear Instruments and Methods in Physics Research Section A: Accelerators, Spectrometers, Detectors and Associated Equipment* **568**, 488-502 (2006).
18. de Loos, M. J. *et al.* Compression of Subrelativistic Space-Charge-Dominated Electron Bunches for Single-Shot Femtosecond Electron Diffraction. *Physical Review Letters* **105**, 264801 (2010).
19. Maxson, J. *et al.* Direct Measurement of Sub-10 fs Relativistic Electron Beams with Ultralow Emittance. *Physical Review Letters* **118**, 154802 (2017).
20. van Oudheusden, T., Nohlmans, J. R., Roelofs, W. S. C., Op't Root, W. P. E. M. & Luiten, O. J. in *Ultrafast Phenomena XVI*. (eds Paul Corkum *et al.*) 938-940 (Springer Berlin Heidelberg).
21. Gao, M., Jiang, Y., Kassier, G. H. & Dwayne Miller, R. J. Single shot time stamping of ultrabright radio frequency compressed electron pulses. *Applied Physics Letters* **103**, 033503 (2013).
22. Kozák, M. *et al.* Optical gating and streaking of free electrons with sub-optical cycle precision. *Nature Communications* **8**, 14342 (2017).
23. Zhou, C. *et al.* Direct mapping of attosecond electron dynamics. *Nature Photonics* **15**, 216-221 (2020).

| | |
|---|---|
| 24 | Zhang, B. *et al.* 1.4‐mJ high energy terahertz radiation from lithium niobates. *Laser & Photonics Reviews* **15**, 2000295 (2021). |
| 25 | Tóth, G., Polónyi, G. & Hebling, J. Tilted pulse front pumping techniques for efficient terahertz pulse generation. *Light: Science & Applications* **12**, 256 (2023). |
| 26 | Jang, D., Sung, J. H., Lee, S. K., Kang, C. & Kim, K. Y. Generation of 0.7 mJ multicycle 15 THz radiation by phase-matched optical rectification in lithium niobate. *Opt Lett* **45**, 3617-3620 (2020). |
| 27 | Baek, I. H. *et al.* Real-time ultrafast oscilloscope with a relativistic electron bunch train. *Nature Communications* **12**, 6851 (2021). |
| 28 | Zhao, L. *et al.* Terahertz Oscilloscope for Recording Time Information of Ultrashort Electron Beams. *Physical Review Letters* **122**, 144801 (2019). |
| 29 | Berden, G. *et al.* Benchmarking of Electro-Optic Monitors for Femtosecond Electron Bunches. *Physical Review Letters* **99**, 164801 (2007). |
| 30 | Cavalieri, A. L. *et al.* Clocking Femtosecond X Rays. *Physical Review Letters* **94**, 114801 (2005). |
| 31 | Bajlekov, S. I. *et al.* Longitudinal electron bunch profile reconstruction by performing phase retrieval on coherent transition radiation spectra. *Physical Review Special Topics - Accelerators and Beams* **16**, 040701 (2013). |
| 32 | LaBerge, M. *et al.* Revealing the three-dimensional structure of microbunched plasma-wakefield-accelerated electron beams. *Nature Photonics* **18**, 952-959 (2024). |
| 33 | Schmidt, B., Lockmann, N. M., Schmüser, P. & Wesch, S. Benchmarking coherent radiation spectroscopy as a tool for high-resolution bunch shape reconstruction at free-electron lasers. *Physical Review Accelerators and Beams* **23**, 062801 (2020). |
| 34 | Gao, M. *et al.* Full characterization of RF compressed femtosecond electron pulses using ponderomotive scattering. *Optics Express* **20**, 12048-12058 (2012). |
| 35 | Gibson, P. J. in *1979 9th European Microwave Conference.* 101-105 (1979). |
| 36 | Rebeiz, G. M. Millimeter-wave and terahertz integrated circuit antennas. *Proceedings of the IEEE* **80**, 1748-1770 (1992). |